\documentstyle[aps,pra,preprint]{revtex}
\begin{document}
\draft
\title{Metastabilities across disorder-broadened first-order transitions} 
\author{P. Chaddah}
\address{Cryogenics and Superconductivity Section,\\
Centre for Advanced Technology, \\ Indore 452013.}
\date{\today }
\maketitle
\begin{abstract}

We  discuss first-order phase  transitions that are broadened by disorder, but still remain first order on the local mesoscopic level. Using vortex-matter as our paradigm, we argue that phase  transitions in general can be broadened by two different causes viz. due to a distribution of the phase transition point (field or temperature) from one region of the sample to another, or due to a distribution in the local magnetic field experienced by different regions of the sample. We show that in the second case there is an apparent ``extrinsic" hysteresis even if the transition is locally second-order. We show that bulk measurements under isothermal conditions can be used to infer whether the broadened transition is first-order at the local (mesoscopic) level.


\end{abstract}

\section{Introduction}

First order (or second order) phase transitions are defined as being accompanied by discontinuities in the first (or higher) derivatives of the free energy with respect to the control variables (temperature T or pressure P) causing the transition. The sample is expected to transform from phase A to phase X on the phase transition line (T$_C$,P$_C$) . This sharp change in the fraction of a particular phase occurs as the (T$_C$,P$_C$) line is crossed, and this results in sharp changes in some physical properties.

Even though the inequality between the free energies of the two phases changes sign on the phase transition line, metastable supercooled/superheated states can persist across a first-order transition till the respective spinodals (called the limit of supercooling or superheating) are reached. These spinodals correspond to the metastable states becoming unstable in that the free energy of the supercooled/superheated state is no longer a local minimum \cite{1,2}. There is still a sharp change in the fraction of a particular phase, but this occurs on a (T$^*$,P$^*$) line while cooling, and on a (T$^{**}$,P$^{**}$) line while heating. The sharp changes in physical properties also occur when the control variables fall on these lines, and a distinct hysteresis is seen while locating the phase transition. This is a characteristic feature of a first-order transition, and has been proposed as a means for identifying the nature of a phase transition \cite{3}. This has to be taken with some caution because hysteresis could also be arising from hindered kinetics \cite{4}, and we shall discuss this possibility in detail. We shall show that the nature of minor hysteresis loops, and the path-dependence (in the space of the two control variables) that is expected for supercooled states \cite{2,5}, can be used to determine the origin of the hysteresis.

\section{Broadening of first order phase transitions.}

\subsection{``Intensive" broadening.}

Imry and Wortis \cite{6} had argued that weak disorder will broaden a first-order transition (or make it `rounded'). Inhomogeneous disorder will cause a distribution of (T$_C$,P$_C$) across the sample so that different regions of the sample change from phase A to phase X at different values of (T$_C$,P$_C$). The (T$_C$,P$_C$) line will now be broadened into a band \cite{7,8}, with the disorder-induced distribution f(T$_C$,P$_C$) dictating the width of the band or of the broadened transition. The extent of broadening does not directly depend on the size of the sample, and we shall refer to this origin of broadening as ``intensive". 

A physical manifestation of this origin has been convincingly demonstrated experimentally by Soibel et al \cite{8} who studied vortex-lattice-melting using a micro-Hall probe. Their measurements showed that ``although the local melting is found to be first-order, a global rounding of the transition is observed; this results from a disorder-induced broad distribution of local melting temperatures, at scales down to the mesoscopic level" \cite{8}. They established that the transition at mesoscopic level was first-order because they could show hysteretic supercooling of microscopic liquid domains.

In this case of a disorder-induced distribution f(T$_C$,P$_C$) across the sample, the fraction F of a particular phase (say phase X) consequently changes smoothly as P (or T) is varied over the width of the (T$_C$,P$_C$) band (see figure 1). Since F changes smoothly, no sharp change in any physical property is expected in any measurement that averages over the entire sample. At this bulk level any phase transition appears like a continuous second order transition. Only if measurements could be done over mesoscopic or microscopic regions, as was done by Soibel et al \cite{8}, then the phase-fraction would be seen to be changing sharply in each region but with different regions undergoing the change at different values of (T,P). Supercooling/superheating would be seen in these microscopic regions (as was seen by Soibel et al) if the transition is first order at the local mesoscopic level, and this results in hysteresis in F. Therefore hysteresis would be seen in some physical properties even in bulk measurements averaging over the entire sample. In the absence of this local supercooling or superheating, F vs H will not show any hysteresis.

It follows that a transition which is intensively broadened will show hysteresis in bulk measurements if it is first order in nature. If the transition in the absence of disorder is continuous, then hysteresis will not be seen because there is no supercooling or superheating. Can we conclude that if a broad transition shows hysteresis then it is first order at the microscopic level? We shall see below that this converse is not true because ``extensively" broadened transitions will show hysteresis even when they are second order in nature.

\subsection{``Extensive" broadening.}

We now discuss a second origin of broadening which we call ``extensive" because the extent of broadening rises as sample size rises. First-order transitions in vortex-matter are expected to be broadened by pinning (or hindered kinetics) of vortices. This hindered kinetics has been understood within Bean's Critical State model (CSM) \cite{9}, and has many interesting manifestations, the most common being M-H hysteresis. What concerns us here is that pinning causes a variation in the magnetic induction {\bf  B} (and thus in the vortex density n) across the sample. The hindered kinetics prevents an equilibration of vortex density as the external field H, which is the analog of pressure P, is varied . Vortex matter in the hard superconductor thus behaves like a material which can sustain an internal pressure gradient. Here one has to distinguish between local pressure (corresponding to {\bf B}(r) or vortex density n(r)) and externally applied pressure (corresponding to applied {\bf H}).  Since various phase transitions occur at different temperatures when at different pressures, different regions of vortex matter in a hard superconductor should undergo a particular phase transition at different temperatures. Similarly, if the temperature is held constant and we are varying the applied magnetic field, then since the density of vortices is position-dependent, different regions of the sample will undergo the vortex matter phase transition at different values of the applied field (but at the same B$_C$ or the same value n$_C$ of local vortex density in each region). The (T$_C$,B$_C$) line remains sharp, but the (T$_C$,H$_C$) line is broadened into a band.

A convincing experimental justification for this idea was provided by the microhall-probe measurements of vortex melting by Zeldov et al \cite{10}. They showed that melting in different regions of the sample occurs at different values of the applied magnetic field, but at the same value of the local field (or vortex density). They showed that this local melting was a sharp transition, while the bulk transition was broad. This idea has been used to explain away the width of transformations that are otherwise expected to be sharp (first order) phase transitions. Specifically, it has been used for explaining the observed width of the vortex melting transition \cite{11}, and of the vortex solid-solid transition at the onset of peak effect \cite{12}. Welp et al \cite{11} and Roy et al \cite{12} argued that because different regions of the sample reach the phase transition line (T$_C$,B$_C$) at different values of the externally applied control variables (T,H), the phase fraction varies continuously. 

As stated in the abstract, we are using vortex matter as a paradigm for studying broadened first-order phase transitions. This is because experimental data exists for both types of broadenings, and the CSM provides an established model for quantifying a discussion on ``extensively" broadened transitions. But one can construct mental pictures of broadening for other first-order transitions as well. Let us consider that water is in a large container consisting of a large number of small containers with impenetrable walls. We shall first assume that these walls are perfect pressure transmitters so that the pressure in each small container is identical. We fill each small container with eutectics of different brines (i.e. different salts in water). The different brines have different freezing points, and the small containers will show first order transitions at different (T,P). The transition in bulk measurements over the big container will show a broadened (step-wise) transition. In the second case, make the walls of the small containers such that they withstand a pressure gradient. If the external pressure is varied, the walls no longer transmit the external pressure faithfully, and the water in each small container experiences a different pressure. Even if each small container is filled with pure water, freezing in each small container occurs at different external (T,P). We would see an extensively broadened transition in the freezing of pure water also! We now go back to vortex matter as our paradigm.

We will now discuss this pinning-induced broadened transition in vortex matter using Bean's CSM \cite{9}. We will obtain the phase fraction F as a function of applied field H and show that it (a) has a width that rises as the sample size rises; and (b) is accompanied by a hysteresis when measurements average over the entire sample. We shall show that this hysteresis is not related to supercooling/superheating, but would be seen even if the transition were continuous on the local scale.

\section{Bean's CSM and ``extensively" broadened transitions.}

We assume that a phase transition to phase X occurs in a region around {\bf r} when {\bf B}(r) crosses B$_C$ . Since {\bf B}(r) is related to the applied {\bf H} through Bean's CSM, it follows that while {\bf B}(r) in the surface region equals B$_C$ when $\mu_0$H = B$_C$ , B(r) at the centre of the sample becomes equal to B$_C$ only when $\mu_0$ H = B$_C$ + H$^*$ if H is being raised, and only when $\mu_0$ H = B$_C$ - H$^*$ if H is being lowered. In the simplest form of Bean's CSM, J$_C$(B) = J$_C$, and H$^*$ = $\mu_0$ J$_C$R, where 2R is the sample dimension perpendicular to the applied field direction \cite{9}. (Further {\bf B} and {\bf H} are scalars for a zero demagnetisation factor sample.) The schematic field profiles B(r) vs r are shown in figure 2. The width of the transition is H$^*$. It increases with increase of sample size R, thus justifying the term "extensively" broadened. The transition which occurs at a particular value B$_C$ of B(r), when seen with applied H as the control variable, now extends from $\mu_0$H = B$_C$ to B$_C$ + H$^*$ when H is being raised, and from B$_C$ - H$^*$ to B$_C$ when H is being lowered. In figure 2 we plot the volume fraction F of the sample having B(r) $>$ B$_C$ in these two cases. This volume fraction is obtained by following standard methods of CSM (see, e.g. references\cite{13,14}) assuming that the sample is an infinite slab parallel to the applied field H. We note a broadening, and hysteresis in volume fraction F having B(r) $>$ B$_C$, as a function of H. We have assumed in figure 2 that the sample shape is of an infinite slab in parallel field which results in a linear variation of the phase fraction; for any other sample shape the fraction would vary non-linearly but hysteresis would still be seen. The extent of broadening and of hysteresis would still be related to H$^*$, as for the slab.

If there is no supercooling or superheating, then phase X exists wherever B(r) $>$ B$_C$, and phase A exists wherever B(r) $<$ B$_C$. In this case F is also the fraction of phase X in the sample. Nowhere in the above discussion have we assumed that the transition at B$_C$ is first order; nonetheless we have ahysteresis in F vs H. The hysteresis is not related to supercooling or superheating, but simply to the local B(r) lagging behind the applied H. Thus, phase transitions that are ``extensively" broadened by disorder would show hysteresis irrespective of their order. The extent of hysteresis also depends linearly on H$^*$ and, for want of a better nomenclature, we term this as ``extrinsic" hysteresis as against ``intrinsic" hysteresis associated with supercooling/superheating. While transitions that are ``intensively" broadened which show hysteresis only if they are first order in nature, extensively broadened transition will always show hysteresis in F. Extensively broadened first-order transitions will show both intrinsic and extrinsic hysteresis.

We have argued that hysteresis is not a sufficient condition to infer that the transition is first order in nature. If one observes hysteresis in a physical property in bulk measurements averaged over the sample, and the transition is broad, can we still check if it is really a first order transition? Are there any distinguishing features for the two origins of hysteresis?

\section{Distinguishing features for the two origins of hysteresis?}

It has been argued earlier that these two types of hysteresis can be distinguished by following different paths in the space of the two control variables; specifically that supercooling/superheating persist farther when only temperature is varied, than when density is varied \cite{2,5}. We expect the opposite path-dependence for hysteresis which is purely kinetic in origin, and have illustrated this using the Critical state model \cite{5}. These tests, of course, can still be used to differentiate intrinsic and extrinsic hysteresis  We present below tests that involve only isothermal variations in applied field (or pressure) and can distinguish whether the hysteresis implies a first-order transition.

\subsection{Minor loops on hysteresis in phase fraction.}

Minor loops (MLs) have been calculated for magnetisation hysteresis in generalised CSM in literature \cite{13}, and B$_r$ is obtained therein under various cyclings of the applied field. Here we are discussing the simple case of field-independent J$_C$ , and we show in figure 3 the field profiles for the cases where field is reversed after being raised to [B$_C$+(H$^*$/2)]. These can be used to calculate F vs H, and obtain minor loops. Figure 3 also shows the phase fractions thus obtained, for three minor loops, started from [B$_C$+(H$^*$/4)], from [B$_C$+(H$^*$/2)], and from [B$_C$+(3H$^*$/4)]. The envelope hysteresis curve is also shown.  We note that MLs drawn from different points on the envelope curve do not merge as long as they are showing finite  hysteresis. Merger can occur only after the hysteresis in the MLs has reduced to zero. This is a specific characteristic feature of ``extrinsic" hysteresis, and this test can be used on the physical property showing hysteresis. If such non-merging MLs are seen, then the hysteresis definitely has an extrinsic component. (If the broadening is only ``intensive" and the hysteresis is intrinsic, then the MLs drawn from different starting points on the envelope curve are expected to merge before the hysteresis reduces to zero. This is depicted in figure 4.)  Non-merging MLs must be seen in the case of extrinsic hysteresis, but do not rule out an intrinsic part; and therefore do not rule out a first-order transition.

\subsection{Effects of pressure (or field) oscillations.}

The effect of pressure oscillations on supercooled states in general, and of magnetic field oscillations on supercooled vortex matter in particular, have been discussed in detail \cite{2,5}. It has been argued that such oscillations would reduce the extent of hysteresis that is intrinsic in origin. The question remains as to what would be the effect of field oscillations on extrinsic hysteresis, within Bean's CSM. If the amplitude of oscillation is reduced slowly to zero, then the phase fraction will change towards the equilibrium value and hysteresis will reduce \cite{13,15}. This behaviour is qualitatively similar to that expected for supercooling/superheating. But if the amplitude of oscillation is kept fixed then, unlike the case of supercooling/superheating, no reduction in hysteresis and no change in phase fraction will be observed \cite{13,15}. This provides a clear method of determining whether there is any supercooling/superheating, because any reduction in hysteresis under oscillations of fixed amplitude implies an intrinsic component to the observed hysteresis.

We note that while non-merging MLs are a clear signature that there is an extrinsic contribution to hysteresis, a reduction in hysteresis under oscillating fields of fixed amplitude is a clear signature that the hysteresis has an intrinsic component and that the broad transition is a first-order transition.

\section{Conclusions.}

Using vortex-matter as our paradigm, we have argued that first-order transitions can be broadened by two different causes viz. due to a distribution of the phase transition field, or due to a distribution in the local magnetic field. We have shown that the latter results in an apparent ``extrinsic" hysteresis even if the transition is locally second-order. We have then shown that bulk measurements under isothermal conditions can be used to definitively infer the origin of the hysteresis, and to infer whether the broadened transition is first-order in nature.

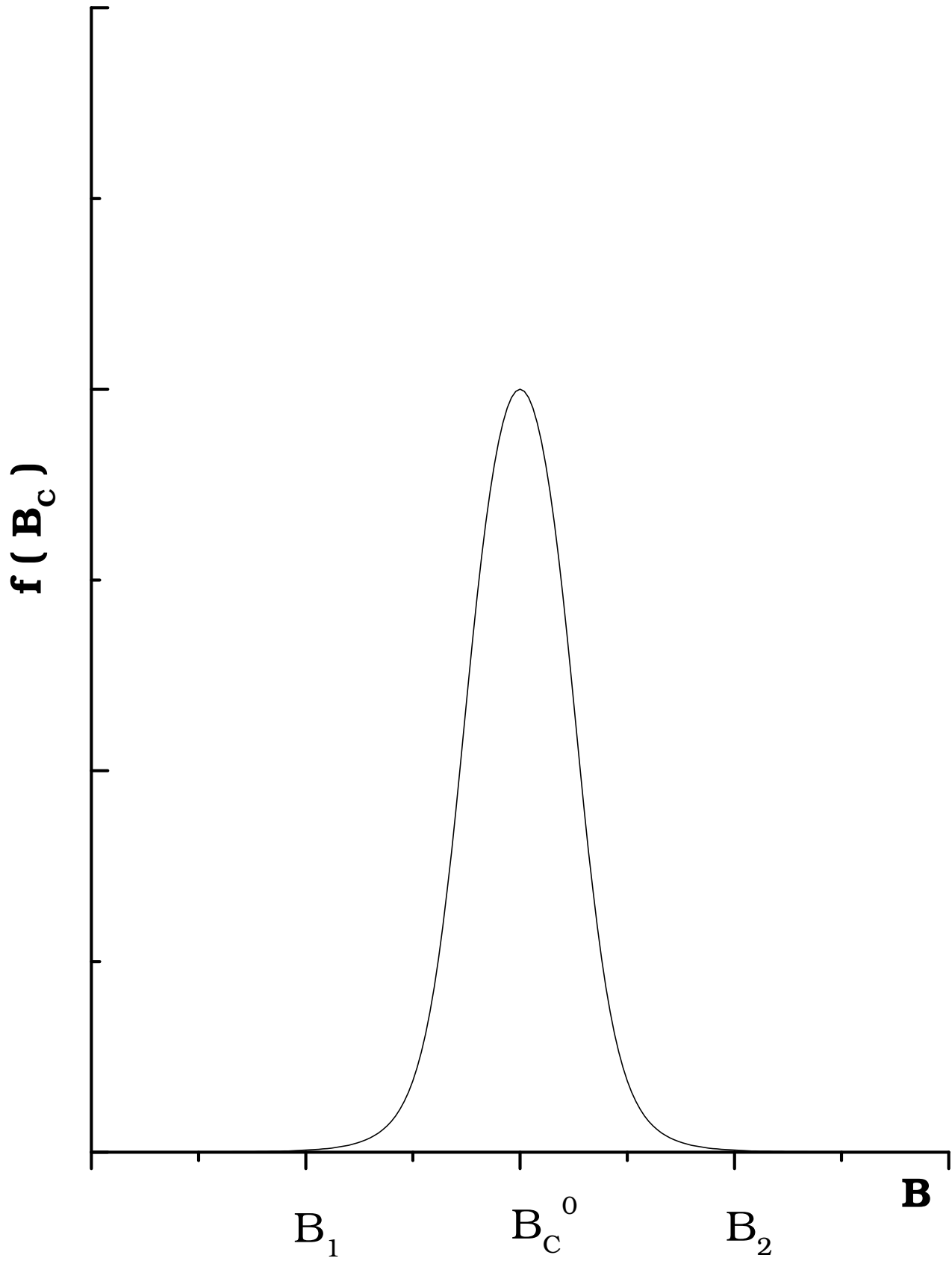
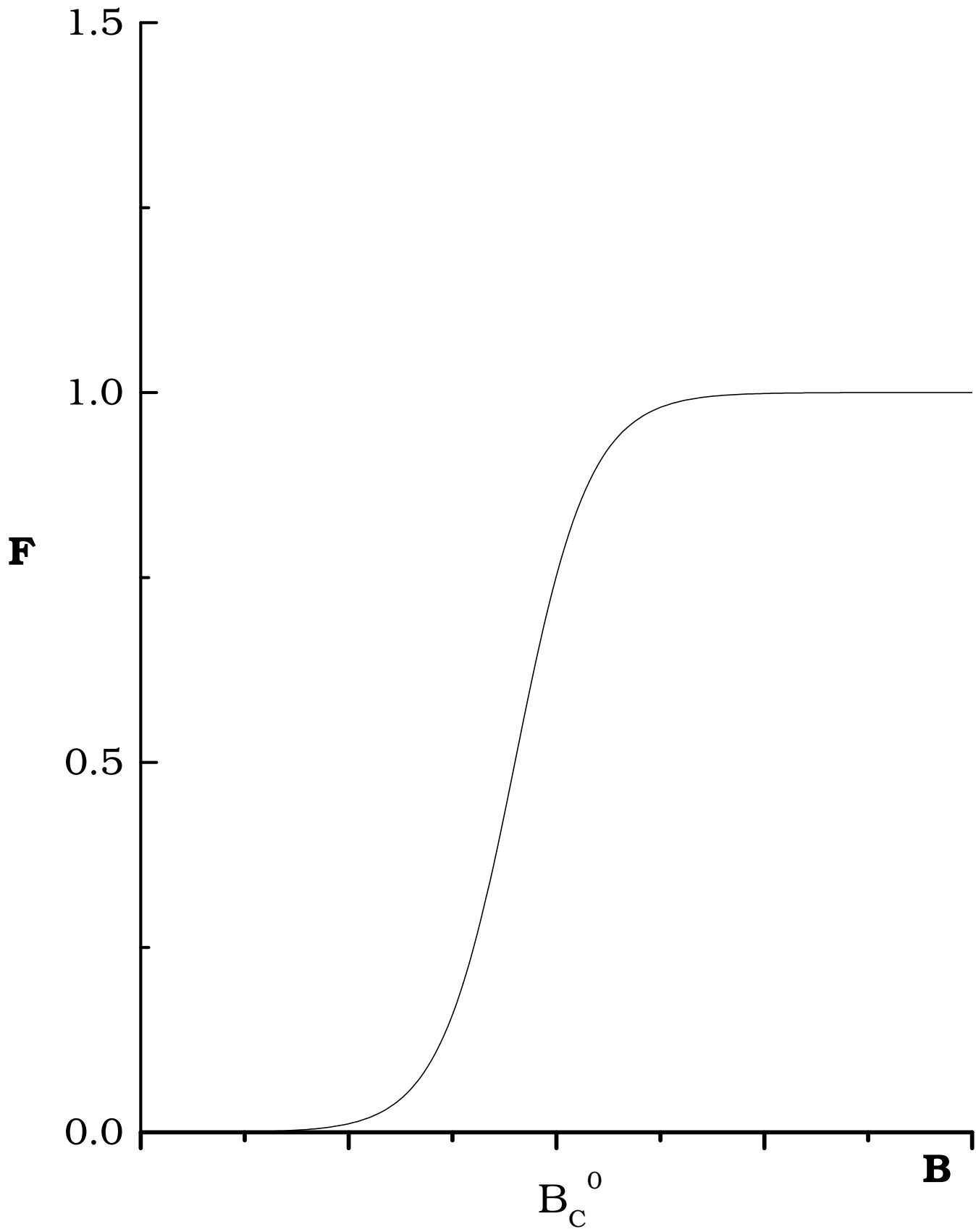
\begin{figure}
\caption{ (a) shows a schematic of the distribution in phase transition fields. (b) shows the resulting smooth rise in the fraction F of the high-field phase. We use field as the variable (instead of pressure) because this draws on the work of Soibel et al (ref [8]) on vortex matter.}
\end{figure}

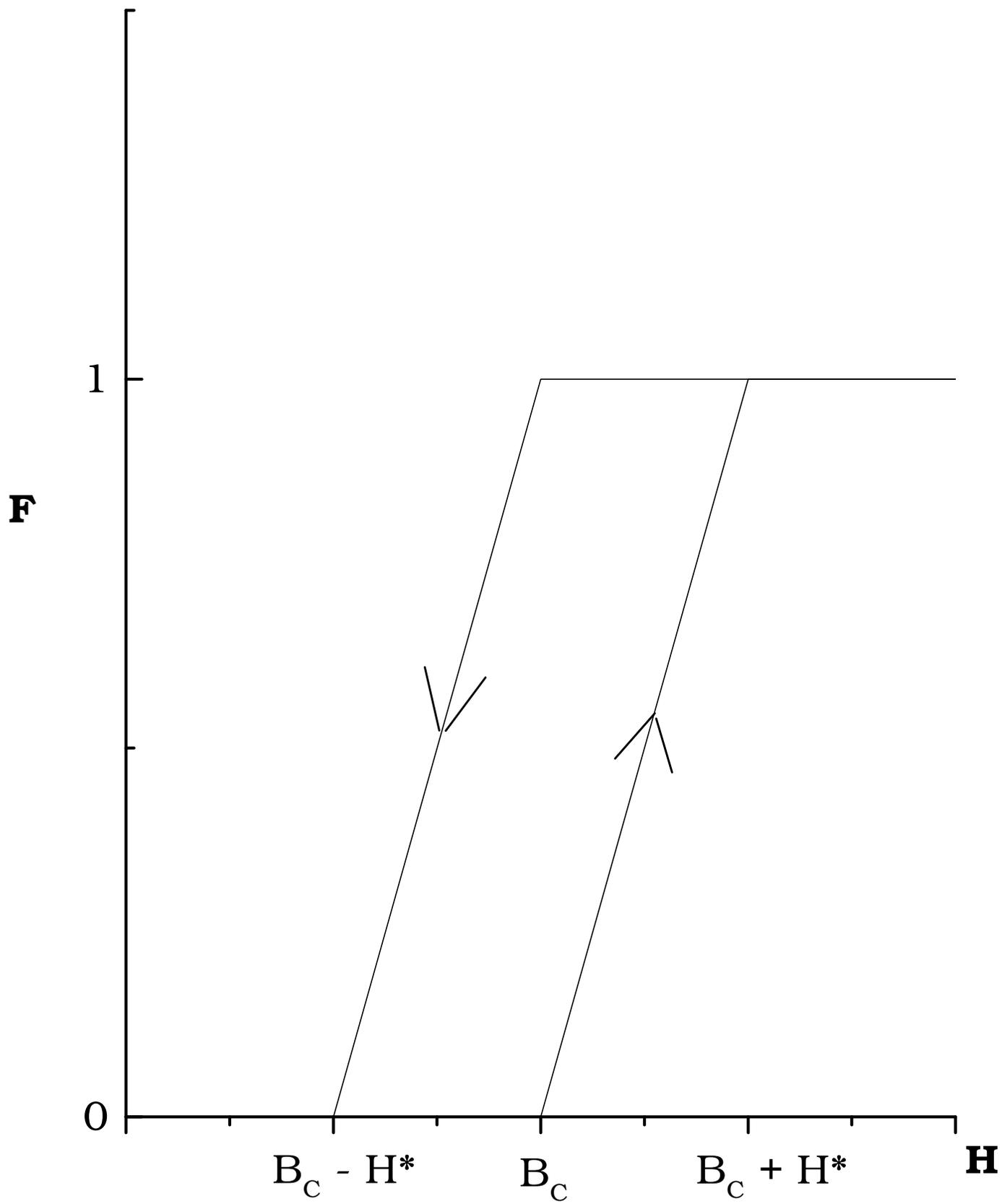
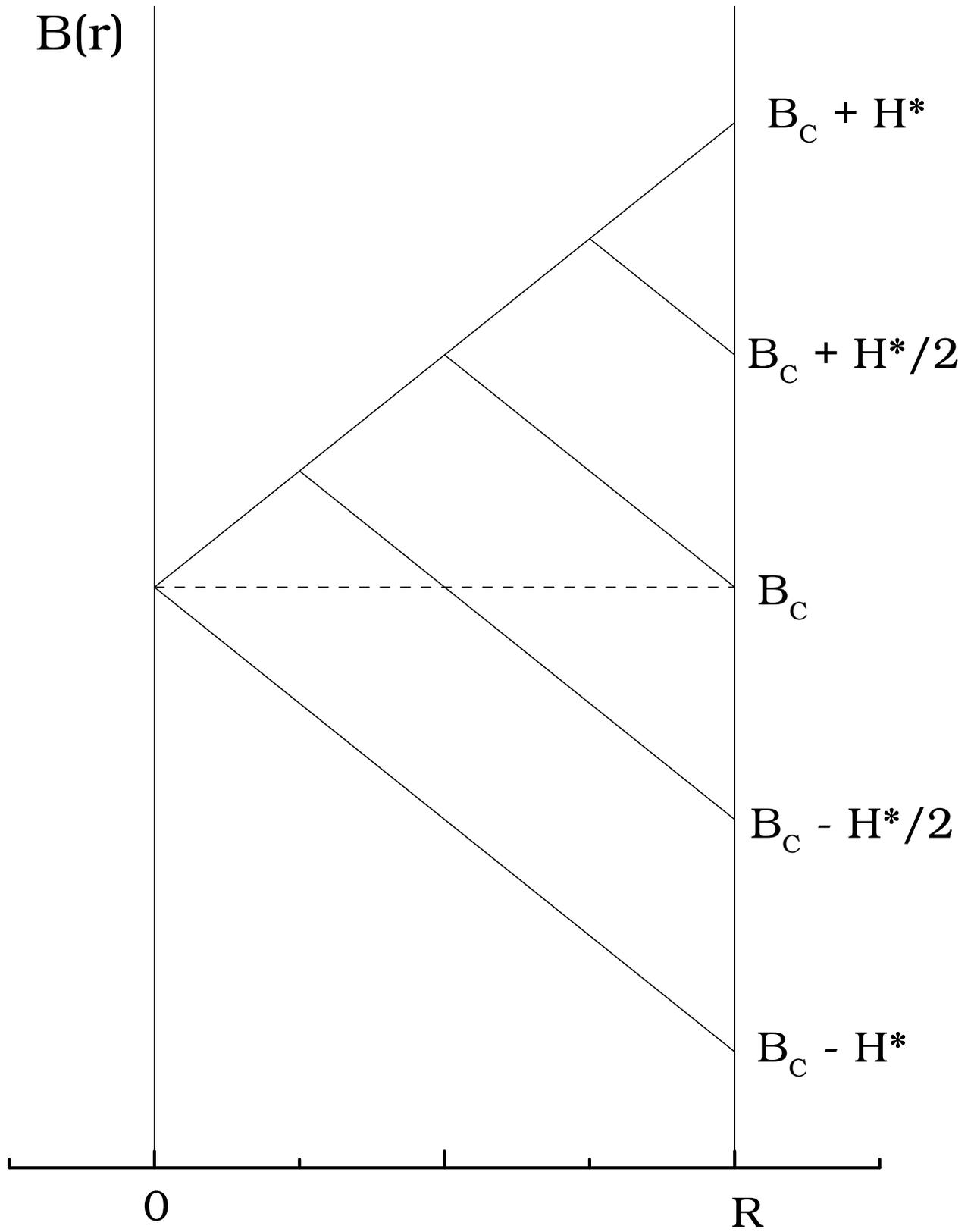
\begin{figure}
\caption{ (a) shows half of the symmetric field distribution, in a slab extending from -R to R, as the external field is raised. The phase fraction rises smoothly, as is shown in (b). The extrinsic hysteresis is seen because B(r) is different on the field-decreasing cycle. The B(r) shown in (c) correspond to the applied field being reduced fron [B$_C$+(H$^*$)] to [B$_C$-(H$^*$)], and give the F vs H envelope for the field-reducing case. }
\end{figure}

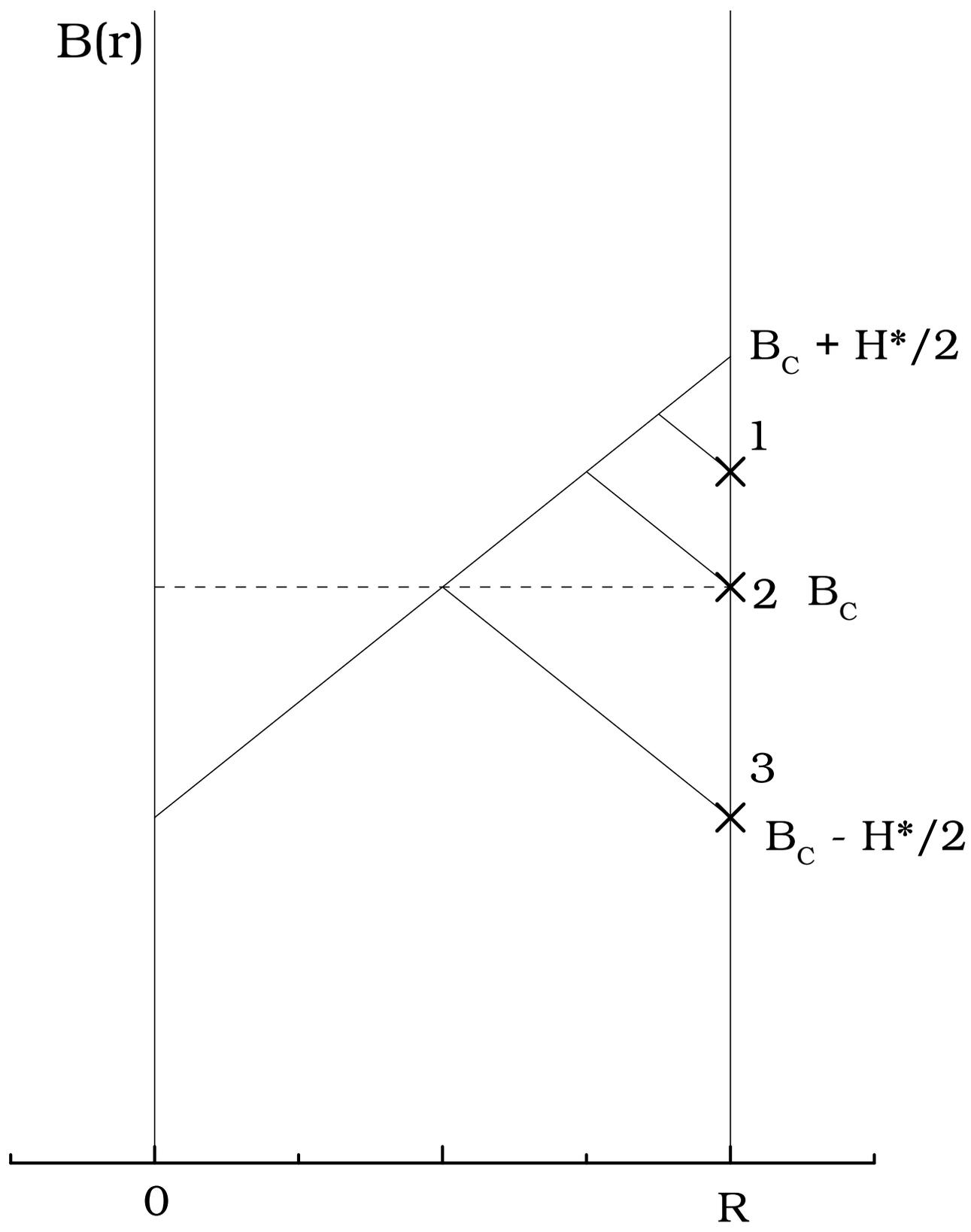
\begin{figure}
\caption{ (a) shows the B(r) obtained when the external field is reversed mid-way from [B$_C$+(H$^*$/2)]. The resulting minor loop in F is shown in (b). The field-profiles for the points marked 1, 2, and 3, are shown in (a). We also show two other minor loops obtained similarly.}
\end{figure}

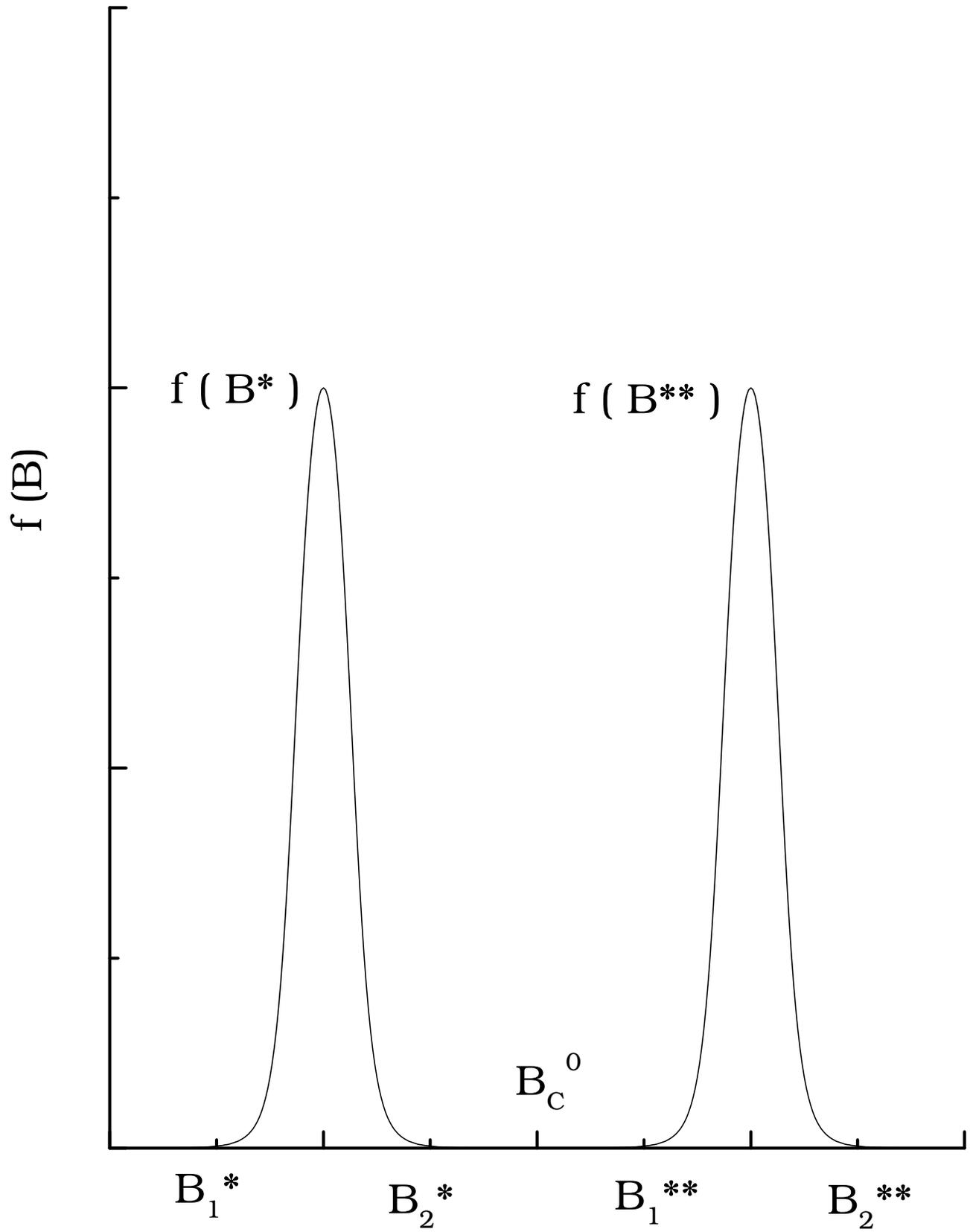
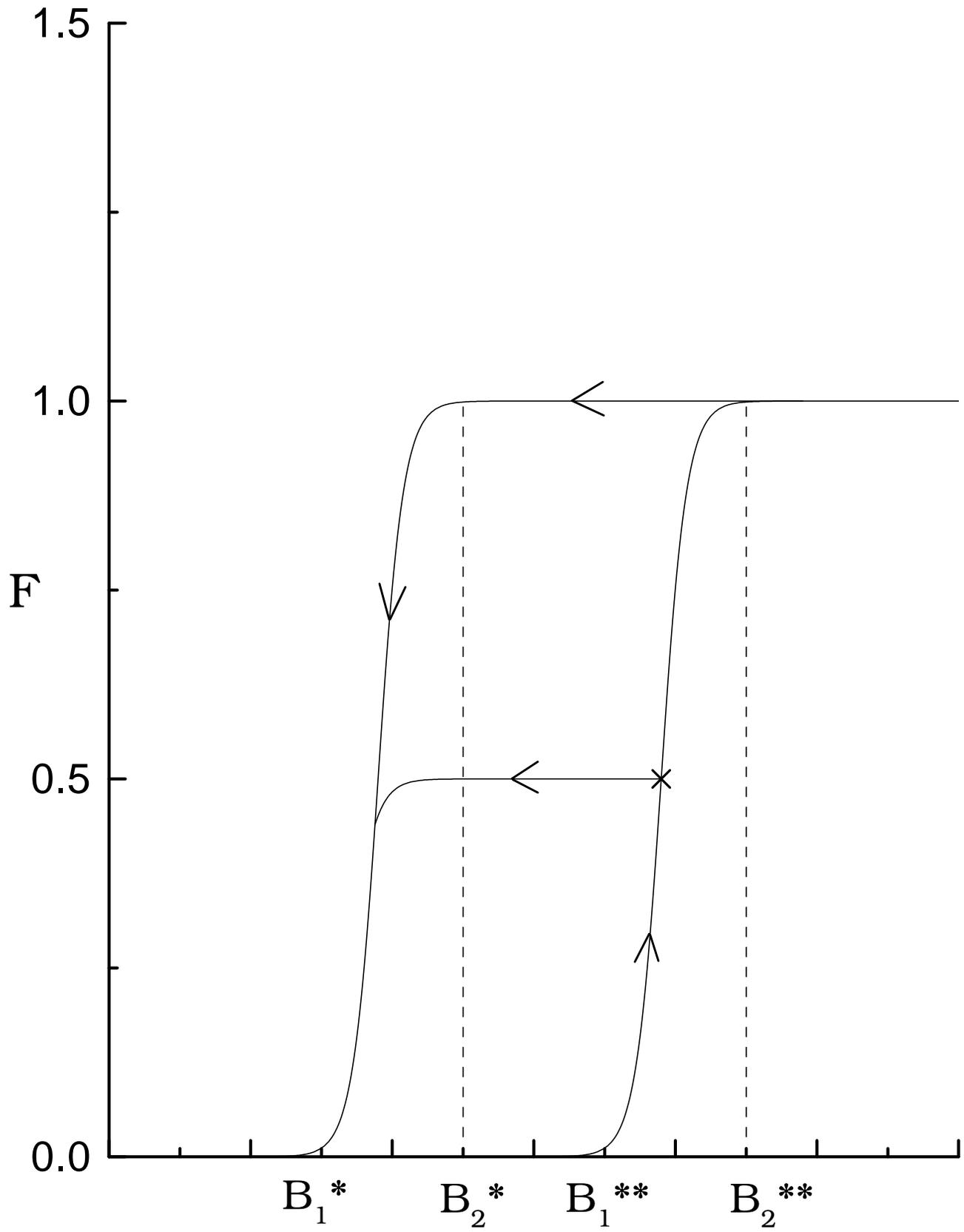
\begin{figure}
\caption{ If there is intrinsic hysteresis, then there will be bands for the limits of supercooling and superheating (ref [7]). These are indicated by the distributions in (a). (b) shows the resulting hysteresis in F, alongwith a minor loop.}
\end{figure}

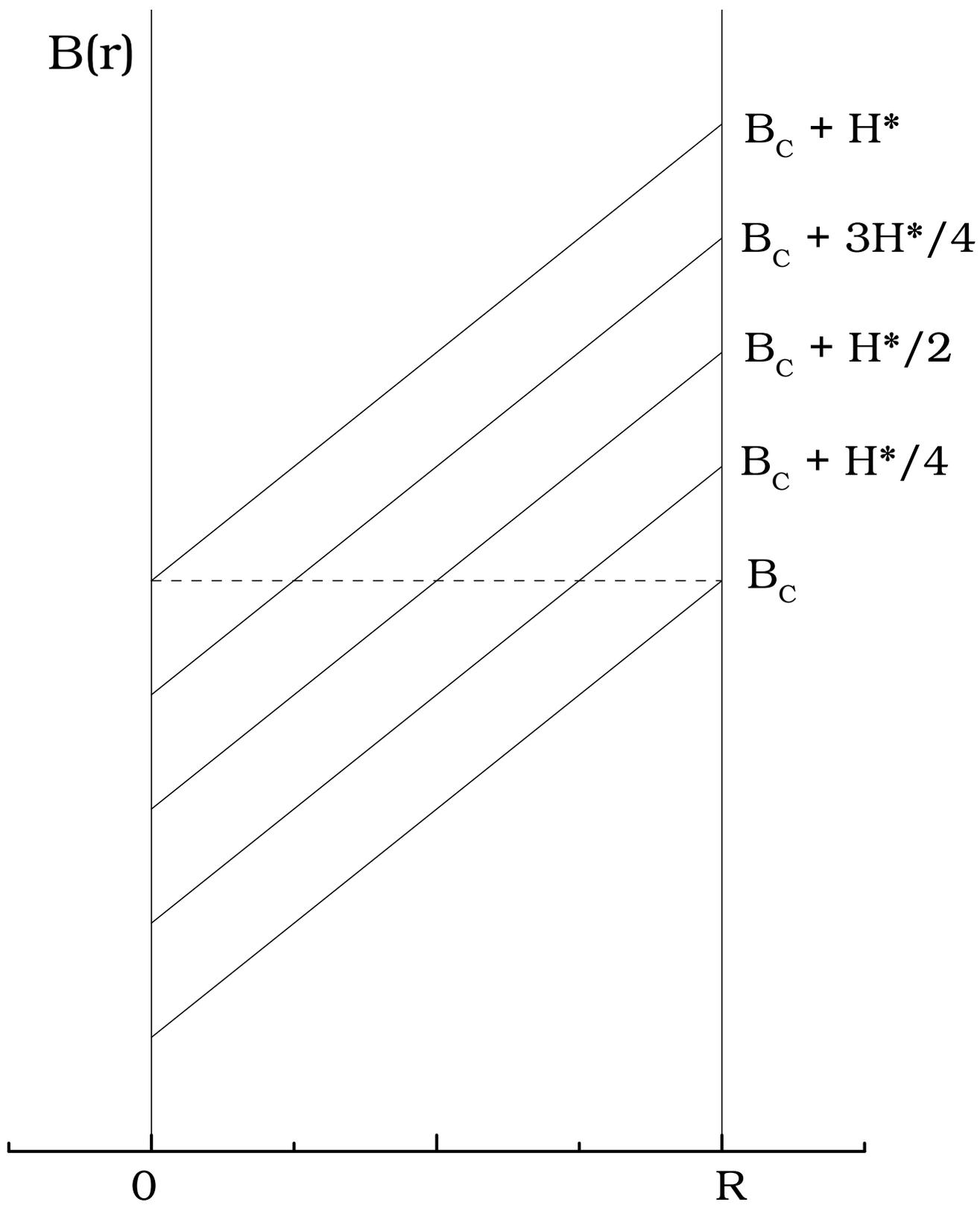
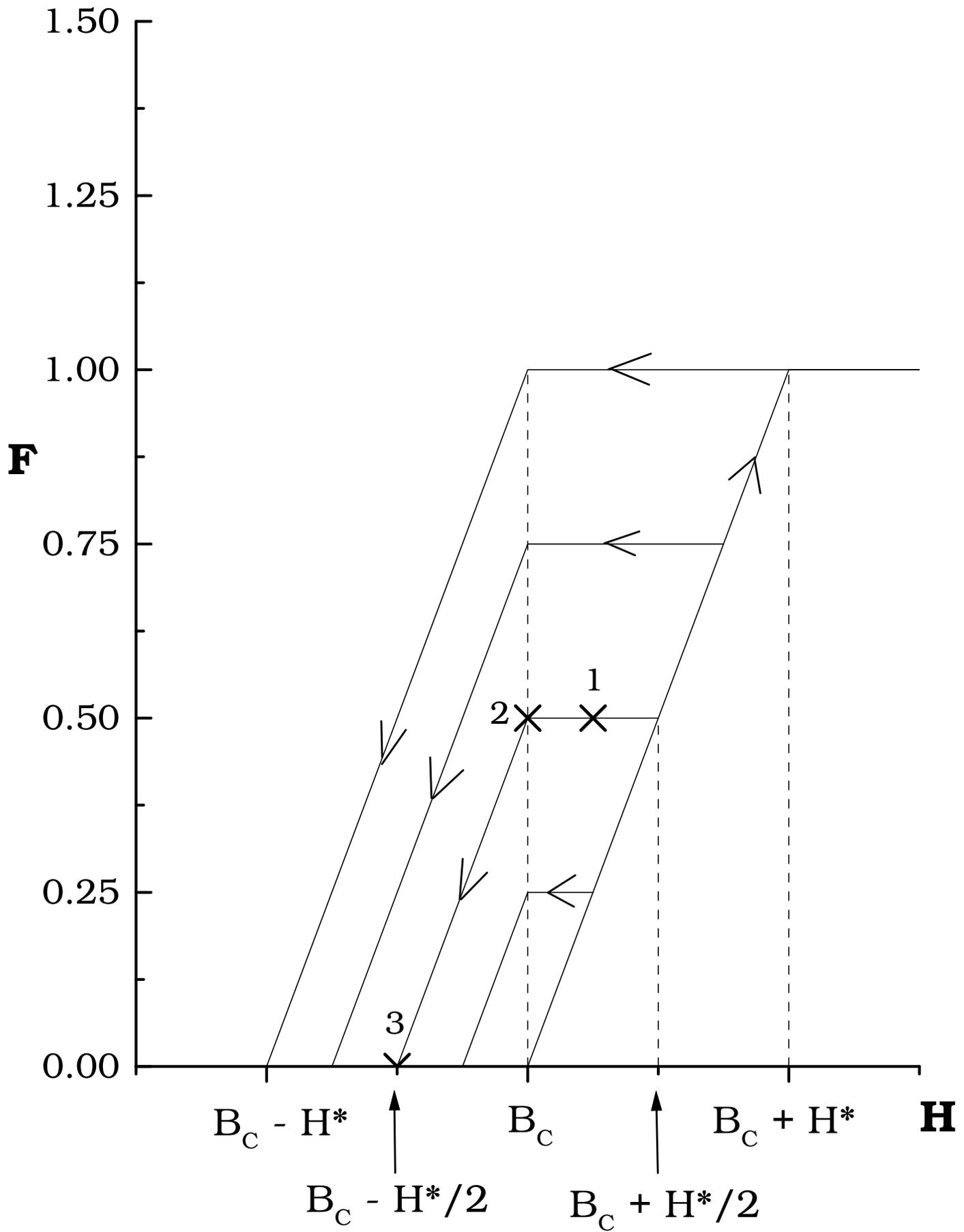
\end{document}